\documentclass[letterpaper,10pt,conference]{ieeeconf}

\IEEEoverridecommandlockouts
\usepackage{amsmath,amssymb,mathrsfs,bm}
\usepackage{caption,subfigure}
\usepackage{graphicx,color,xspace}
\usepackage{algorithm,url}

\usepackage[noend]{algpseudocode}
\makeatletter
\def\BState{\State\hskip-\ALG@thistlm}
\makeatother

\usepackage{dsfont}
\usepackage{tikz}

\usepackage{accents}

\newcommand{\oprocendsymbol}{\hbox{$\bullet$}}
\newcommand{\oprocend}{\relax\ifmmode\else\unskip\hfill\fi\oprocendsymbol}

\newcommand{\longthmtitle}[1]{\mbox{}\textup{(#1):}}

\newtheorem{theorem}{Theorem}[section]

\newtheorem{remark}[theorem]{Remark}

\newtheorem{proposition}[theorem]{Proposition}

\renewcommand{\baselinestretch}{1}

\newcommand{\bmse}{{\rm{BMSE}}}
\newcommand{\cbmse}{{\rm{CBMSE}}}
\newcommand{\bzb}{{\rm{BZB}}}

\newcommand{\sk}{s_k}
\newcommand{\xjk}{x_{j, k}}
\newcommand{\sjk}{s_{j, k}}
\newcommand{\gk}{{\sensingpol_k}}
\newcommand{\gktx}{{G_k^\text{\scalebox{0.9}{Tx}}}}
\newcommand{\gkrx}{{G_k^\text{\scalebox{0.9}{Rx}}}}
\newcommand{\sensingpol}{G}
\newcommand{\gall}[1]{\sensingpol^{(#1)}}
\newcommand{\gon}[1]{\sensingpol_{#1}}
\newcommand{\gstark}{{\sensingpol_k^*}}

\newcommand{\njk}{n_{j,k}}

\newcommand{\signoise}{\sigma}

\newcommand{\arx}{{a}^\text{\scalebox{0.9}{Rx}}}
\newcommand{\atx}{{a}^\text{\scalebox{0.9}{Tx}}}

\newcommand{\mksteer}{{m}_k}
\newcommand{\mksimo}{{m}_{k}^\text{\scalebox{0.7}{SIMO}}}
\newcommand{\mkmimo}{{m}_{k}^\text{\scalebox{0.7}{MIMO}}}

\newcommand{\gammavec}{\bm{\gamma}}

\newcommand{\nact}{{I}}
\newcommand{\nactrx}{{I^\text{\scalebox{0.9}{Rx}}}}
\newcommand{\nacttx}{{I^\text{\scalebox{0.9}{Tx}}}}
\newcommand{\nrx}{N}
\newcommand{\mtx}{M}

\newcommand{\tpulse}{T}

\newcommand{\ubar}[1]{\underline{#1}}

\newcommand{\idtx}{d^\text{\scalebox{0.9}{Tx}}}
\newcommand{\idrx}{d^\text{\scalebox{0.9}{Rx}}}

\newcommand{\dtx}{\ubar{d}^\text{\scalebox{0.9}{Tx}}}
\newcommand{\drx}{\ubar{d}^\text{\scalebox{0.9}{Rx}}}

\newcommand{\dvirtarg}[1]{\ubar{d}^\text{\scalebox{0.9}{Virt}}_{#1}}
\newcommand{\dtdmvirt}{\ubar{d}^\text{\scalebox{0.9}{TDM}}}
\newcommand{\dtxm}[1]{\idtx_{#1}}
\newcommand{\drxn}[1]{\idrx_{#1}}

\newcommand{\wwb}{{\rm{WWB}}}

\newcommand{\etak}{\eta_k}

\newcommand{\expecd}[1]{\mathbb{E}_{#1}}

\newcommand{\parg}[2]{p(#1|#2)}

\newcommand{\quotposhat}[1]{\hat{\phi}_{#1}}
\newcommand{\quotpos}[1]{\phi_{#1}}
\newcommand{\ppos}[1]{p_{#1}}
\newcommand{\intlikelihoods}[1]{D_{#1}}
\newcommand{\intlikelihoodsmimo}[1]{D_{#1}^\text{\scalebox{0.7}{MIMO}}}
\newcommand{\intlikelihoodssimo}[1]{D_{#1}^\text{\scalebox{0.7}{SIMO}}}

\newcommand{\utheta}{\theta}
\newcommand{\fd}{f_D}

\newcommand{\pargenh}{\hat{\pargen}}
\newcommand{\pargen}{\theta}

\newcommand{\xall}[1]{X^{(#1)}}
\newcommand{\xon}[1]{X^{#1}}

\newcommand{\repart}[1]{\operatorname{Re}\{#1\}}

\newcommand{\tp}{^{\text{\scalebox{0.8}{$\top$}}}}
\newcommand{\her}{^{\text{\scalebox{0.8}{H}}}}
\newcommand{\defin}{:=}

\newcommand{\until}[1]{\{1,\dots,#1\}}

\newcommand{\setdef}[2]{\{#1 \, : \, #2\}}

\newcommand{\real}{{\mathbb{R}}}

\newcommand{\zeroonematrix}[2]{\{0,1\}^{#1\times #2}}

\newcommand{\complex}{{\mathbb{C}}}

\newcommand{\norm}[1]{\|#1\|_{\text{\scalebox{1}{$2$}}}}

\newcommand{\ones}{\mathds{1}}
\newcommand{\identity}{\mathrm{I}}

\begin{document}




\title{Adaptive channel selection for DOA estimation in  MIMO radar
}

\author{David Mateos-N\'u\~nez 
\qquad Mar\'ia A. Gonz\'alez-Huici
\qquad Renato Simoni
\qquad Stefan Br\"uggenwirth
\thanks{The
    authors are with the department of Cognitive Radar at Fraunhofer FHR, Wachtberg, Germany, 
    {\tt \small
   \{david.mateos-nunez, maria.gonzalez, renato.simoni, stefan.brueggenwirth\}@fhr.fraunhofer.de}.
%
}
}

\maketitle

\begin{abstract}
We present adaptive strategies for antenna selection for Direction of Arrival (DoA) estimation of a far-field source using TDM MIMO radar with linear arrays. Our treatment is formulated within a general adaptive sensing framework that uses one-step ahead predictions of the Bayesian MSE using a parametric family of Weiss-Weinstein bounds that depend on previous measurements. 
We compare in simulations our strategy with adaptive policies that optimize the Bobrovsky-Zaka\"i bound and the Expected Cram\'er-Rao bound, and show the performance for different levels of measurement noise.
\end{abstract}

\begin{keywords}
	Adaptive Sensing, Antenna Selection, Array Processing, Weiss-Weinstein Bound, Bayesian Filtering, Direction of Arrival (DOA), MIMO, Cognitive Radar.
\end{keywords}

\section{Introduction}
Recent advances in millimeter-wave radar circuits
make possible low-cost and compact multi-channel radar systems that can be controlled by software. This motivates the design of signal generation and processing algorithms that attempt to maximize the information extracted from the scene, in what is considered the basis of the perception-action cycle of a 
%
cognitive radar architecture~\cite{SH:06,JE-SB:15}.
 



Algorithms for adaptive transmission
typically employ a prediction of the conditional Bayesian mean-square error (BMSE) given previous observations. In the category of adaptive strategies that attempt to optimize one-step ahead predictions, recent works optimize parameters such as the pulse repetition frequency (PRF) in Pulse-Doppler radar in a joint framework for detection and tracking~\cite{KLB-JTJ-GES-CJB-MR:15a, KLB-CJB-GES-JTJ-MR:15b}, or the transmitted signal autocorrelation matrix in MIMO radar for DoA estimation~\cite{WH-JT-RS:13,NS-JT-HM:15}, using, respectively, the conditional Bayesian Cram\'er-Rao bound (BCRB) and the Reuven-Messer bound (RMB)~\cite{IR-HM:97}.
In the category of algorithms that consider the consequences of actions based on some long-term reward, the work~\cite{AC-FH:15} schedules measurements in a tracking scenario where the target is temporarily occluded, and~\cite{MF-SH:14} optimizes waveform parameters using planning and reinforcement learning.

Few works have considered these approaches for adaptive antenna selection for Direction of Arrival (DoA) estimation. Accuracy of angular estimation improves with the length of the antenna array, and thus with the number of antenna elements that need to be adequately spaced to avoid ambiguity due to aliasing. 
Bigger apertures demand more Tx and Rx modules 
(and hence a higher system cost) 
and more data to be processed in real time. 
%
To overcome these constraints, the works~\cite{OI-JT-IB:15,JT-OI-IB:16} study adaptive receiver selection algorithms for far-field DoA and SNR estimation with SIMO linear arrays based on optimization of the Bobrovsky-Zaka\"i Bound (BZB), which provides better one-step ahead predictions than the Expected CRB (ECRB). The latter is not sensitive to sidelobe level
but is related to the average mainlobe width of the array~\cite{UB-RLM:03}, 
and selects the receivers that yield biggest aperture regardless of previous measurements~\cite{OI-JT-IB:15}.
Alternatively, the Weiss-Weinstein bound~\cite{AW-EW:85} is computationally more expensive but predicts more accurately the contribution to the Mean-Square-Error (MSE) of sidelobe ambiguity at low SNR~\cite{HN-HLVT:94,AR-PF-PL-CDR-AN:08}.
%

We extend the work of~\cite{OI-JT-IB:15,JT-OI-IB:16} to transmitter and receiver selection for DoA estimation in Time Domain Multiplexing (TDM) MIMO radar with linear arrays and propose a general algorithm for adaptive sensing using the Weiss-Weinstein bound. 
 Using a particle filter~\cite{AD-SG-CA:98} to incorporate sequentially the information from measurements into the belief distribution of the unknown parameter, we
 construct the conditional WWB 
 (along with the BZB and the ECRB, for reference),
 that lower bounds the achievable MSE. This requires a double optimization procedure, first over the so-called test-points, to evaluate the tightest bound, and then over candidate sensing parameters. The resulting strategies are illustrated in simulations where we compare the performance of channel selection based on the WWB, the BZB, and the ECRB.
 %
%

The rest of the paper is organized as follows: Section~\ref{sec:adaptive-sensing-framework}
proposes a general framework for adaptive sensing based on one-step ahead predictions of the MSE.  The general strategy is then particularized in Section~\ref{sec:particularization-adaptive-channel-selection} to MIMO channel selection for DoA estimation. Finally, Section~\ref{sec:conclussions} presents our conclusions and ideas for future work.

\section{Adaptive sensing via Weiss-Weinstein bound}\label{sec:adaptive-sensing-framework}

In this section we present the general strategy to optimize sensing parameters based on a prediction of the MSE. First we introduce the WWB, then we connect it to the conditional BMSE, and finally we describe the computation of the conditional WWB involved in our algorithm.


\subsection{Preliminaries on the Weiss-Weinstein bound}
The WWB 
provides a lower bound on the BMSE of any estimator and thus gives an indication of the achievable estimation performance. Namely, the expected error of any estimator $\pargenh(x)$, over possible pairs of observations~$x$ and one-dimensional parameter values~$\theta$ modeled with probability distribution $p(x,\pargen)$, is bounded as follows,
\begin{align}\label{eq:ineq-bmse-wwb-general}
\expecd{p(x,\pargen)}
\big[(\pargenh(x)-\pargen)^2\big] \ge\wwb(s,h) ,
\end{align}
where $\wwb(s,h)$ is a member of the parametric family of bounds~\cite{AW-EW:85},~\cite[eq. 76]{AR-PF-PL-CDR-AN:08} given by
\begin{align}\label{eq:def-wwb-parametric}
\wwb(s,h)\defin
\frac{h^2 \eta(s,h)^2}{\eta(2s,h)+\eta(2-2s,-h)-2\eta(s,2h)} ,
\end{align}
where~$\eta$ is the moment generating function~\cite[pp. 337, pp. 65]{HLVT-KLB:13} 
defined as\footnote{The usual convention is to consider the logarithm, but in our presentation we choose this notation for convenience.}
\begin{align}\label{eq:def-eta-general}
\eta(\alpha,\beta)\defin&\,\int_\Theta \int_\Omega 
\frac{p^\alpha(x, \utheta+\beta)}{p^{\alpha-1}(x, \utheta)} dx d\utheta
\notag
\\
=&\,\int_\Theta \Big(\int_\Omega 
\frac{p^\alpha(x|\utheta+\beta)}{p^{\alpha-1}(x| \utheta)} dx \Big)
\frac{p^\alpha(\utheta+\beta)}{p^{\alpha-1}( \utheta)} d\utheta,
\end{align}
where $p(x|\utheta)$ is the probability, or \textit{likelihood}, of the observation~$x\in\Omega\subseteq\real^n$ given the parameter value~$\utheta\in\Theta\defin\setdef{\theta\in\real}{p(\utheta)>0}$, and~$p(\utheta)$ is the \textit{prior} probability distribution of~$\utheta$, which is considered a modeling choice. The value of the so-called \textit{test-point}~$h\in(0,\infty)$, and the additional degree of freedom~$s\in(0,1)$, determine the bound on the BMSE, the tightest bound being obtained as
$\wwb\defin\sup_{s,h}\wwb(s,h)$ .
%
(For further generalizations we refer the reader to~\cite{HLVT-KLB:13}.)
%
The BZB can be obtained from~\eqref{eq:def-wwb-parametric} in the limit cases~$s=1$ or $s=0$, 
\begin{align}\label{def-bzb-parametric}
\bzb(h)=\wwb(s=1,h)=\frac{h^2 }{\eta(2,h)-1} .
\end{align}
%
%
The BCRB~\cite[pp. 72]{HLVT-KLB:13} is in turn a particular case of~\eqref{def-bzb-parametric}, under suitable assumptions on the differentiability of~$p(\utheta)$, in the limit as $h\to 0$.
In the next section we present the connection between the WWB described here and the conditional BMSE, relevant for our adaptive strategies.

\subsection{Conditional BMSE and adaptive sensing}

Consider an estimation task where a sequence of observations~$\xall{k-1}\defin (\xon{1},\dots,\xon{k-1})$ of an unknown parameter~$\theta$ is obtained using a sequence of sensing parameters~$\gall{k}\defin (\gon{1},\dots,\gon{k})$, in a suitable domain, according to an observation model with 
joint probability distribution
$\parg{\xall{k},\pargen}{\gall{k}}$.
An adaptive sensing strategy or policy can be defined in general by a probability distribution over sensing parameters given previous measurements, $\parg{\gon{k}}{\xall{k-1},\gall{k-1}}$.
In this work, the proposed strategies are evaluated with respect to the BMSE, which is defined, for any estimator~$\pargenh\equiv\pargenh(\xall{k},\gall{k})$, as the following integration over observations and realizations of the parameter,
\begin{align*}
&\;\bmse(\pargenh,\gall{k})\defin
\expecd{\parg{\xall{k},\pargen}{\gall{k}}}
\big[(\pargenh-\pargen)^2\big]
\\
=&\; \expecd{\parg{\xall{k-1}}{\gall{k}}}\Big[\expecd{\parg{\xon{k},\pargen}{\xall{k-1},\gall{k}}}\big[(\pargenh-\pargen)^2\big]\Big] .
\end{align*}
Following~\cite{WH-JT-RS:13} and~\cite{JT-OI-IB:16}, we 
consider the inner expectation above, called conditional BMSE (CBMSE),
\begin{align*}
\cbmse(\pargenh, \xall{k-1},\gall{k})
\defin 
\expecd{\parg{\xon{k}, \pargen}{\xall{k-1}, \gall{k}}}\big[(\pargenh-\pargen)^2\big] ,
\end{align*}
as an optimization metric for adaptive algorithms that attempt to find, at each step~$k\ge 1$, a policy~$\gk$ that minimizes the BMSE given any sequences of previous sensing policies $\gall{k-1}$ and historical observations~$\xall{k-1}$. 
Such metric is usually impossible to compute explicitly, but can be lower-bounded in a similar fashion as the BMSE
in relation~\eqref{eq:ineq-bmse-wwb-general}. 
Motivated by this observation, we define the parametric family of conditional WWBs, denoted by
 $\wwb(s,h;\xall{k-1},\gall{k})$, as in~\eqref{eq:def-wwb-parametric},
where in the definition~\eqref{eq:def-eta-general} we use the 
 likelihood function~$\parg{\xon{k}}{\pargen,\gall{k} }$ and replace the prior distribution by the \textit{posterior} $\ppos{k-1}(\theta)\defin\parg{\pargen}{\xall{k-1},\gall{k-1}}$. The moment generating function in~\eqref{eq:def-eta-general} becomes then
\begin{align}\label{eq:def-eta-conditional}
\etak(\alpha,\beta)
\defin
&\,\int_\Theta \int_\Omega 
\frac{\parg{x}{\utheta+\beta,\gall{k}}^\alpha}
{\parg{x}{\utheta,\gall{k}}^{\alpha-1}} dx
\frac{\ppos{k-1}(\utheta+\beta)^\alpha}
{\ppos{k-1}(\utheta)^{\alpha-1}} d\utheta,
\end{align}
where $\ppos{0}(\pargen)\defin p(\theta)$ is the prior probability.
(Note that the domain of integration in~\eqref{eq:def-eta-conditional} is such that~$\ppos{k-1}(\pargen)>0$.)
\begin{proposition}\longthmtitle{Conditional WWB and CBMSE}\label{prop:conditional-wwb}
Consider the observation 
model $\parg{\xall{k},\pargen}{\gall{k}}$
under the following two assumptions,~i)~$\xon{k}$ and $\xall{k-1}$ are conditionally independent given~$\theta$ and~$\gall{k}$, i.e.,
$\parg{\xon{k},\xall{k-1}}{\pargen,\gall{k}} 
= 
\parg{\xon{k}}{\pargen,\gall{k} }\parg{\xall{k-1}}{\pargen,\gall{k-1} }$,
%
%
and~ii)~$\parg{\xon{k}}{\pargen,\gall{k} }=\parg{\xon{k}}{\pargen,\gon{k} }$. Then 
\begin{align*}
\cbmse(\pargenh, \xall{k-1},\gall{k}) \ge \wwb(s,h;\xall{k-1},\gall{k}) .
\end{align*}
\end{proposition}
%
%
(The proof is standard and is omitted for lack of space.)
%


Motivated by the above result, we define adaptive strategies that select at step~$k$ the sensing policy~$\gk$ based on knowledge from previous measurements~$\xall{k-1}$ and previous sensing policies~$\gall{k-1}$, as the solution of
\begin{align}\label{eq:policy-optimization-wwb-framework}
\gstark\in\arg\min_\gk \;\sup_{\substack{s\in (0, 1)\\ h\in(0,\infty)}}  \wwb(s,h\,;\xall{k-1},\gall{k}) .
\end{align}
In general, the inner optimization problem in~\eqref{eq:policy-optimization-wwb-framework} over test-points is nonconvex, requiring methods for global optimization such as simulated annealing~\cite{SK-CDG-MPV:83},
and the outer optimization over sensing policies can be discrete. 
In the next section we explain how to construct the parametric family of conditional bounds~$\wwb(s,h;\xall{k-1},\gall{k})$ using the likelihood function, the sequence of measurements~$\xall{k-1}$, and previous sensing policies~$\gall{k-1}$. 


%

\subsection{Computation of the conditional WWB}

To evaluate~$\wwb(s,h;\xall{k-1},\gall{k})$ we first re-write the moment generating function~\eqref{eq:def-eta-conditional}, consistently with the notation in~\cite{JT-OI-IB:16}, as
\begin{align}\label{eq:eta-posterior}
\etak(\alpha,\beta)=&\,
\int_\Theta \intlikelihoods{k}(\theta,\alpha,\beta)\, \quotpos{k-1}(\theta,\alpha,\beta)\,\ppos{k-1}(\pargen)  d\theta
\notag
\\
=&\,\expecd{\ppos{k-1}(\pargen) }\big[\intlikelihoods{k}(\theta,\alpha,\beta)\, \quotpos{k-1}(\theta,\alpha,\beta)
\big] ,
\end{align}
where $\intlikelihoods{k}(\theta,\alpha,\beta)$ contains the observation model,
\begin{align*}
\intlikelihoods{k}(\theta,\alpha,\beta)\defin\int_\Omega 
\frac{\parg{x}{\pargen+\beta,\gk}^\alpha}{\parg{x}{\pargen,\gk}^{\alpha-1}} dx ,
\end{align*}
and $\quotpos{k-1}(\theta, \alpha,\beta)$ contains the posterior distribution,
\begin{align}\label{eq:quotient-priors}
&\, \quotpos{k-1}(\theta, \alpha,\beta)\defin\Big(\frac{\ppos{k-1}(\pargen+\beta)}{\ppos{k-1}(\pargen)}\Big)^\alpha .
\end{align}

The computation of~$\quotpos{k-1}(\theta, \alpha,\beta)$ requires special attention. Again by Bayes Law and using the assumptions of Proposition~\ref{prop:conditional-wwb}, we can express the posterior probability as follows,
\begin{align}\label{eq:posterior-factorization}
\ppos{k-1}(\theta)\defin &\, \parg{\pargen}{\xall{k-1},\gall{k-1}}
\notag
\\
= &\, \frac{\parg{\pargen}{\gall{k-1}}}{\parg{\xall{k-1}}{\gall{k-1}}}\prod_{m=1}^{k-1} \parg{\xon{m}}{\pargen,\gon{m}} .
\end{align}
Next we make an observation that connects the iterative computation of the posterior in~\eqref{eq:posterior-factorization} with the approximation of the expectation in~\eqref{eq:eta-posterior}. 

\begin{figure}[bth]
	\centering 
	{\includegraphics[width=0.99\linewidth]{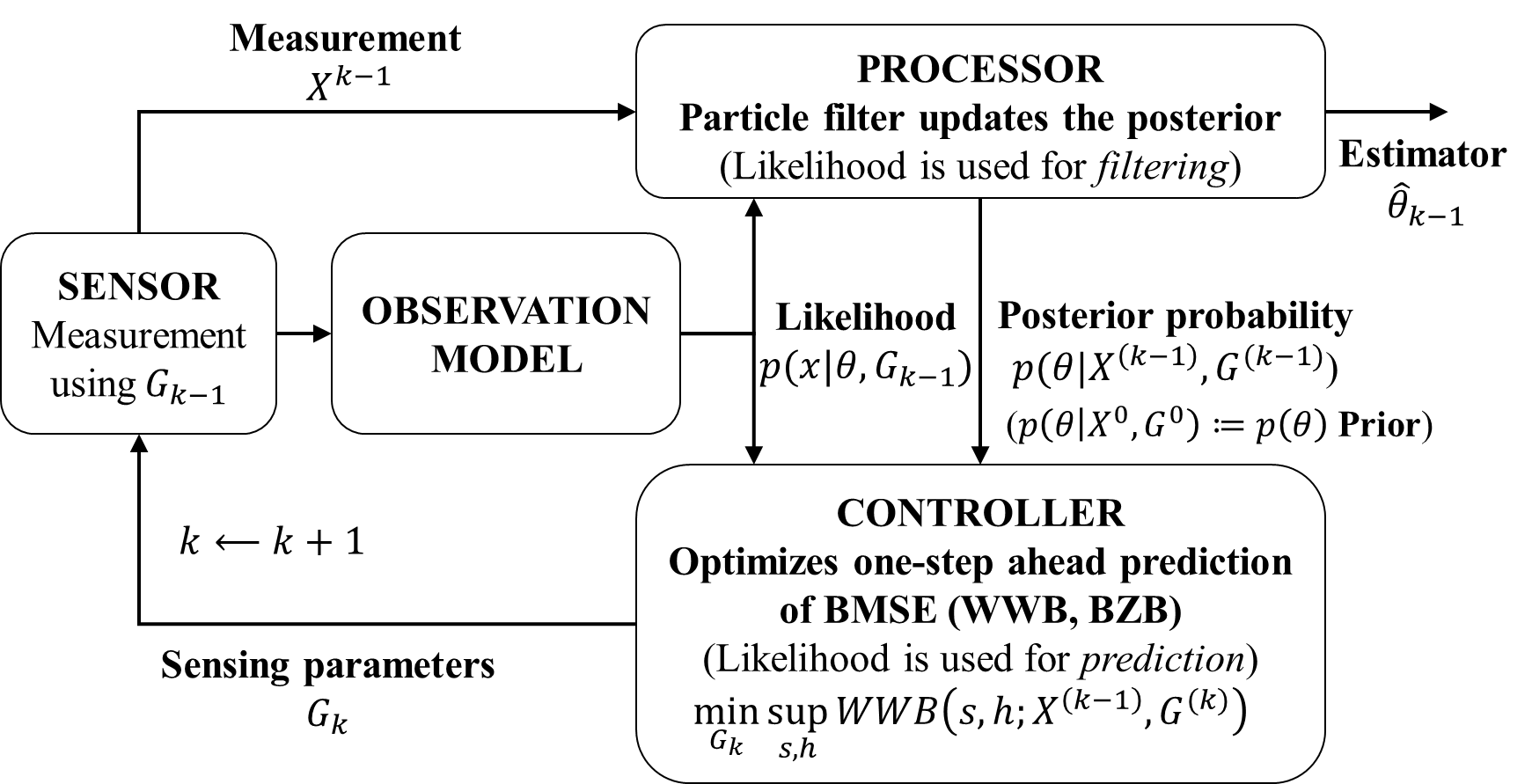}}
	\caption{Basic diagram of adaptive sensing based on WWB.}\label{fig:adaptive-policy-diagram}
\end{figure}

\begin{remark}\longthmtitle{Computation of the expectation~\eqref{eq:eta-posterior} and dependence of measurements in adaptive algorithms}\label{re:posterior-depenednce-measurements}
	Suppose that the sensing parameters are chosen randomly without using previous data, so that $\parg{\pargen}{\gall{k-1}}=\frac{\parg{\gall{k-1}}{\pargen}}{p(\gall{k-1})}p(\pargen)=p(\pargen)$. 
	Under this approximation,\footnote{To our knowledge, the study of filtering performance in scenarios of dependent measurements 
		due to greedy adaptation of the sensing policies is absent in the literature.} the expectation in~\eqref{eq:eta-posterior} can be approximated numerically via Monte Carlo integration leveraging two facts: i) the quotient in~\eqref{eq:quotient-priors} 
	can be factorized, in view of~\eqref{eq:posterior-factorization}, as
	\begin{align*}
	\quotposhat{k-1}(\theta,\alpha,\beta)
	\defin \Big(\frac{p(\pargen+\beta)}{p(\pargen)}
	\frac{\prod_{m=1}^{k-1} \parg{\xon{m}}{\pargen + \beta,\gon{m}}}{\prod_{m=1}^{k-1} \parg{\xon{m}}{\pargen,\gon{m}}}\Big)^\alpha ,
	\end{align*}
	and ii) one can sample from the posterior using Monte Carlo methods such as particle filters.\footnote{For $1$-dimensional cases, integration using the empirical PDF of the particles might be more efficient than Monte Carlo integration. 
		We use this approach for DoA estimation with and without the approximation given by fitting the posterior by a Gaussian, which  lowers the computational cost. 
	}
	%
	 \oprocend
\end{remark}

In summary, measurements depend on the sensing policies, as prescribed by the likelihood. The likelihood serves two purposes, see Fig.~\ref{fig:adaptive-policy-diagram}: i)~filtering in the processor, where particles, i.e., guesses of the parameter, are re-sampled according to which ones make the measurements more likely, and ii)~prediction in the controller, where the WWB is computed integrating the joint distribution that combines the likelihood of possible observations and the current posterior.


\section{Adaptive channel selection}\label{sec:particularization-adaptive-channel-selection}
In this section, we particularize the strategy for adaptive sensing in Section~\ref{sec:adaptive-sensing-framework} to the problem of adaptive channel selection for DoA estimation with MIMO linear arrays.

\subsection{Problem statement}\label{subsec:data-model-ch-selection}

Here we consider the problem of angle of arrival estimation of a single far-field point target. For this, we use a linear array of~$\nact$ omnidirectional antennas, with observation model
\begin{align}\label{eq:data-model}
\xjk = \mksteer(\utheta) \sjk + \njk,
\end{align}
where $\xjk\in\complex^\nact$ is the observation at snapshot~$j\in\until{J}$ in step~$k\in\{1,2,\dots\}$, $\mksteer(\utheta)\in\complex^\nact$ is the steering vector for the 
unknown electronic azimuth $\utheta\defin\sin(\phi)$, where $\phi\in(-\pi,\pi)$ is the azimuth or direction of arrival, $\sjk\in\complex$ is the target signal 
(which here we assume is known), and $\njk$ is the noise, modeled by independent and identically distributed zero-mean complex Gaussians with real and imaginary parts also independent with covariance~$\sigma\identity_\nact$ (i.e., a multiple of the identity matrix).
In SIMO radar (i.e., a single transmitter and multiple receivers), $\mksteer(\utheta)$ corresponds to the receive steering vector $\arx(\utheta)\in\complex^\nrx$, 
which is defined, for a far-field source and $\nrx$ receivers located at positions $\drx\defin [\drxn{1}, \cdots, \drxn{\nrx}]\in\real^\nrx$, as
$\arx(\utheta)\defin e^{j k_0 \drx \utheta} ,$
where~$k_0=2\pi/\lambda$ is the wavenumber and $\lambda$ is the received wavelength.
To incorporate into the model the selection of  Rx elements, we define the \textit{receive switching matrix} $\gkrx\in\zeroonematrix{\nactrx}{N}$, for a total of~$\nactrx$ active receivers, such that the $i$th row contains a nonzero element only in column $n_i$, and each column has at most a nonzero element. 
The \textit{switched receive steering vector} is then defined as
\begin{align}\label{eq:def-switched-receive}
\mksimo(\utheta)\defin\gkrx\arx(\utheta)=&\,
\begin{bmatrix}
e^{j k_0 \drxn{n_1} \utheta}, \cdots, e^{j k_0 \drxn{n_\nactrx} \utheta} \end{bmatrix}\tp
\notag
\\
=&\; e^{j k_0 \gkrx\drx \utheta}.
\end{align}
Similarly, for DoA estimation using MIMO arrays, we define the \textit{switched TDM MIMO steering vector} as
\begin{align}\label{eq:def-switched-transmit-MIMO}
\mkmimo(\utheta)\defin(\gammavec(\fd)
\odot(\gktx\atx(\utheta)))\otimes(\gkrx\arx(\utheta)) , 
\end{align}
where $\atx(\utheta)\defin e^{j k_0 \dtx\utheta}\in\complex^\mtx$ is the transmit steering vector for $\mtx$ transmitters located at positions $\dtx\defin[\dtxm{1}, \cdots, \dtxm{\mtx}]$; the \textit{transmit switching matrix} $\gktx\in\zeroonematrix{\nacttx}{\mtx}$, for~$\nacttx$ active transmitters, is such that the $i$th row contains only a nonzero element in column $m_i$ (and each column has at most a nonzero element); and
 $\gammavec(\fd)\defin e^{j 2\pi  \tpulse [1,\cdots,\nacttx]\fd}\in\complex^\nacttx$ contains the Doppler frequency shift $\fd\in\real$ (that we assume is known here), typical in a TDM scheme. The latter term results from the sequence of pulses from the active transmitters with inter-pulse duration~$T>0$. With this notation, \eqref{eq:def-switched-transmit-MIMO} can be written as
 \begin{align}\label{eq:def-switched-transmit-rewritten}
 &\,\mkmimo(\utheta)
= 
 \notag
 \begin{bmatrix}
 e^{j 2\pi \fd \tpulse 1} e^{j k_0 \dtxm{m_1} \utheta}, \cdots, e^{j 2\pi \fd \tpulse \nacttx} e^{j k_0 \dtxm{m_\nacttx} \utheta} 
 \end{bmatrix}\tp
 \\
 &\,
 \otimes
\begin{bmatrix}
e^{j k_0 \drxn{n_1} \utheta}, \cdots, e^{j k_0 \drxn{n_\nactrx} \utheta} \end{bmatrix}\tp
=   
e^{j (k_0 \dvirtarg{k} \utheta + \dtdmvirt \fd)} 
 , 
 \end{align}
 where  $\dtdmvirt\defin 2\pi \tpulse\, [1,\cdots,\nacttx]\otimes\ones_\nactrx$, and
 \begin{align*}
 \dvirtarg{k}\defin
 \ones_\nacttx \otimes (\gkrx\drx)
 + (\gktx \dtx) \otimes \ones_\nactrx .
 \end{align*}
The goal is to choose a total of~$\nact=\nacttx+\nactrx$ active transmitters and receivers, specified by $\gk=\{\gktx,\gkrx\}$ at each step $k\in\{1,2,\dots\}$, that help extract the maximum amount of information about the angle of arrival according to~\eqref{eq:policy-optimization-wwb-framework}. The only part of the adaptive sensing strategy of Section~\ref{sec:adaptive-sensing-framework} that needs to be particularized is the likelihood function, which naturally depends on the observation model above, cf.~Fig.~\ref{fig:adaptive-policy-diagram}. Using the corresponding likelihood function for DoA estimation in SIMO and MIMO radar, in the next section we construct the WWB associated to these problems.

\subsection{Conditional WWB for DoA estimation}

To apply the general strategy of Section~\ref{sec:adaptive-sensing-framework} to the problem of antenna selection, we need to use the likelihood function associated to the observation model~\eqref{eq:data-model}, see Fig.\ref{fig:adaptive-policy-diagram}.
The
likelihood function of~$J$ snapshots $\xon{k}=[ x_{1,k},\dots, x_{J,k}]$, given $\utheta$ and sensing parameters~$\gk=\{\gktx,\gkrx\}$, is distributed as a product of complex Gaussian distributions
because snapshots are assumed independent, i.e.,
\begin{align*}
\parg{\xon{k}}{\utheta,\gk}
=\prod_{j=1}^J
 \Big(\frac{1}{(\pi\signoise^2)^\nact}
e^{-\tfrac{1}{\signoise^2}
	\norm{x_{j,k}-\mksteer(\utheta)\sjk}^2
} \Big).
\end{align*}
From the computation in~\cite[eq. (137)]{AR-PF-PL-CDR-AN:08}, one has
\begin{align*}
\intlikelihoods{k}(\utheta,\alpha,\beta)=  &\,\prod_{j=1}^J\int_{\real^{2\nact}}
\frac{p^\alpha(x_{j,k}| \utheta+\beta)}{p^{\alpha-1}(x_{j,k}|\utheta)} dx_{j,k}
\notag
\\
=&\,e^{\sk^2 \frac{\alpha(\alpha-1)}{\signoise^2}\norm{\mksteer(\utheta+\beta)-\mksteer(\utheta)}^2},
\end{align*}
where $\sk^2\defin\sum_{j=1}^J |\sjk|^2$.
(Note that the model with unknown stochastic target signals, called \textit{unconditional}, requires a different calculation, cf.~\cite{JT-OI-IB:16,DTV-AR-RB-SM:11}.)
In the SIMO case, using the definition~\eqref{eq:def-switched-receive}, 
we get
	\begin{align*}
&\,\norm{\mksimo(\utheta\!+\!\beta)-\mksimo(\utheta)}^2
=
\norm{\mksimo(\utheta\!+\!\beta)}^2+\norm{\mksimo(\utheta)}^2
\\
&\!-2\repart{\mksimo(\utheta\!+\!\beta)\her\mksimo(\utheta)}
=2\nactrx
-2\repart{\sum_{i=1}^{\nactrx}
	\mksimo(\beta)_i} ,
\end{align*}
which is related to the \textit{ambiguity surface} (cf.~\cite[pp. 269, eq. 4.229]{HLVT-KLB:13}) for the selected receivers. Therefore, 
\begin{align*}
\intlikelihoodssimo{k}(\alpha,\beta)
\defin
e^{\frac{\alpha(\alpha-1)\sk^2}{\signoise^2}\big(2\nactrx-2\sum_{i=1}^\nactrx\cos( k_0 (\gkrx\drx)_i \beta)\big)} .
\end{align*}
Similarly, for the MIMO case, using~\eqref{eq:def-switched-transmit-rewritten}, we obtain
\begin{align*}
\intlikelihoodsmimo{k}(\alpha,\beta)
\defin
e^{\frac{\alpha(\alpha-1)\sk^2}{\signoise^2}
	\big(2\nacttx\nactrx-2\sum_{i=1}^{\nacttx\nactrx}\cos( k_0 (\dvirtarg{k})_i \beta)\big)} .
\end{align*}
%
%
Equipped with the functions~$\intlikelihoodssimo{k}(\alpha,\beta)$ and $\intlikelihoodsmimo{k}(\alpha,\beta)$ (which incidentally do not depend on $\theta$), the parametric family of conditional bounds~$\wwb(s,h;\xall{k-1},\gall{k})$ can be expressed in terms of~\eqref{eq:eta-posterior} according to~\eqref{eq:def-wwb-parametric}. Note that the posterior can be approximated following Remark~\ref{re:posterior-depenednce-measurements}. We can then evaluate candidate sets of channels specified by $\gk=\{\gktx,\gkrx\}$, and select the optimal ones according to~\eqref{eq:policy-optimization-wwb-framework}. 
Next we present simulations with synthetic measurements.


\begin{figure}[bth]
	\begin{center}
	{\includegraphics[width=0.99\linewidth]{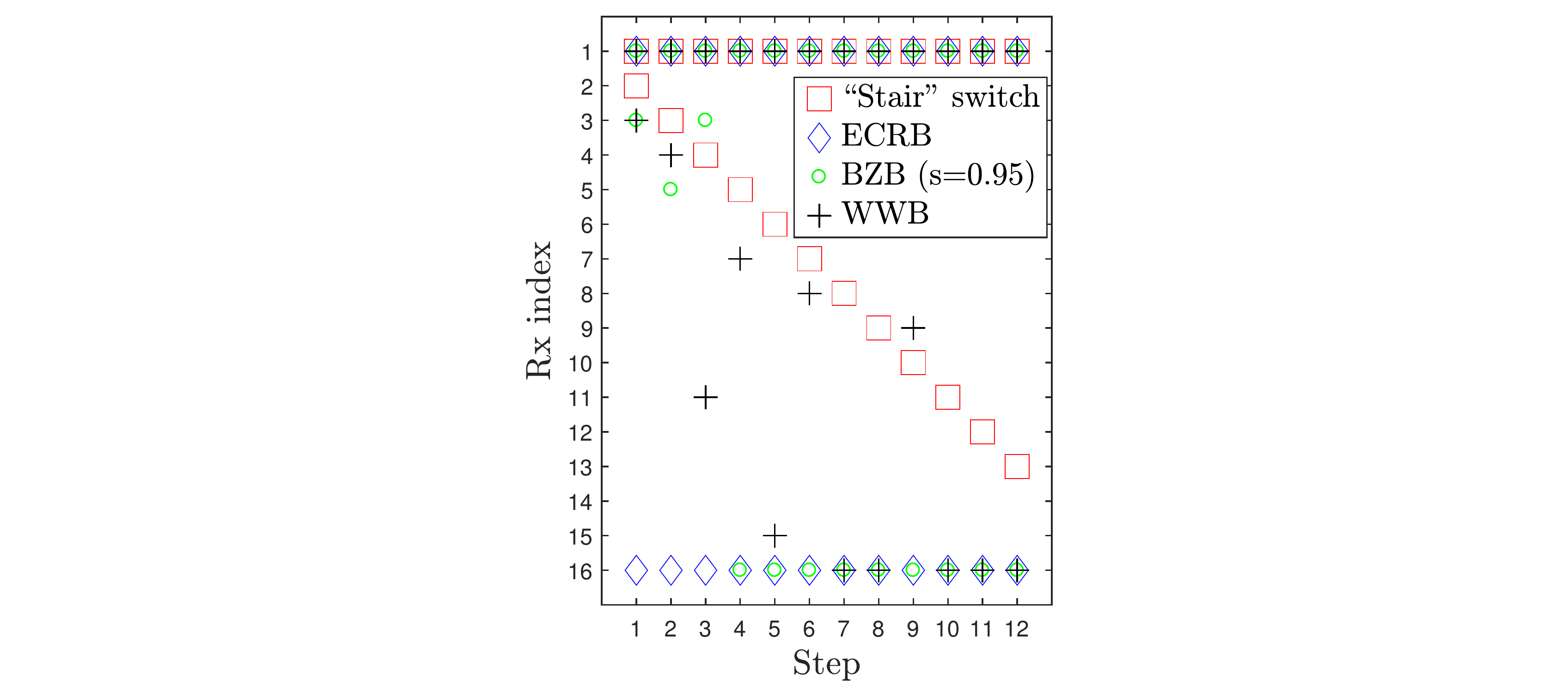}}
	\caption{Optimal channel choices in a typical execution in the SIMO case, with SNR $= -5$, where Rx $1$ is always fixed.}
	\label{fig:channel-activations-SIMO}
	\end{center}
\end{figure}

\begin{figure}[bth]
	\begin{center}
	{\includegraphics[width=0.99\linewidth]{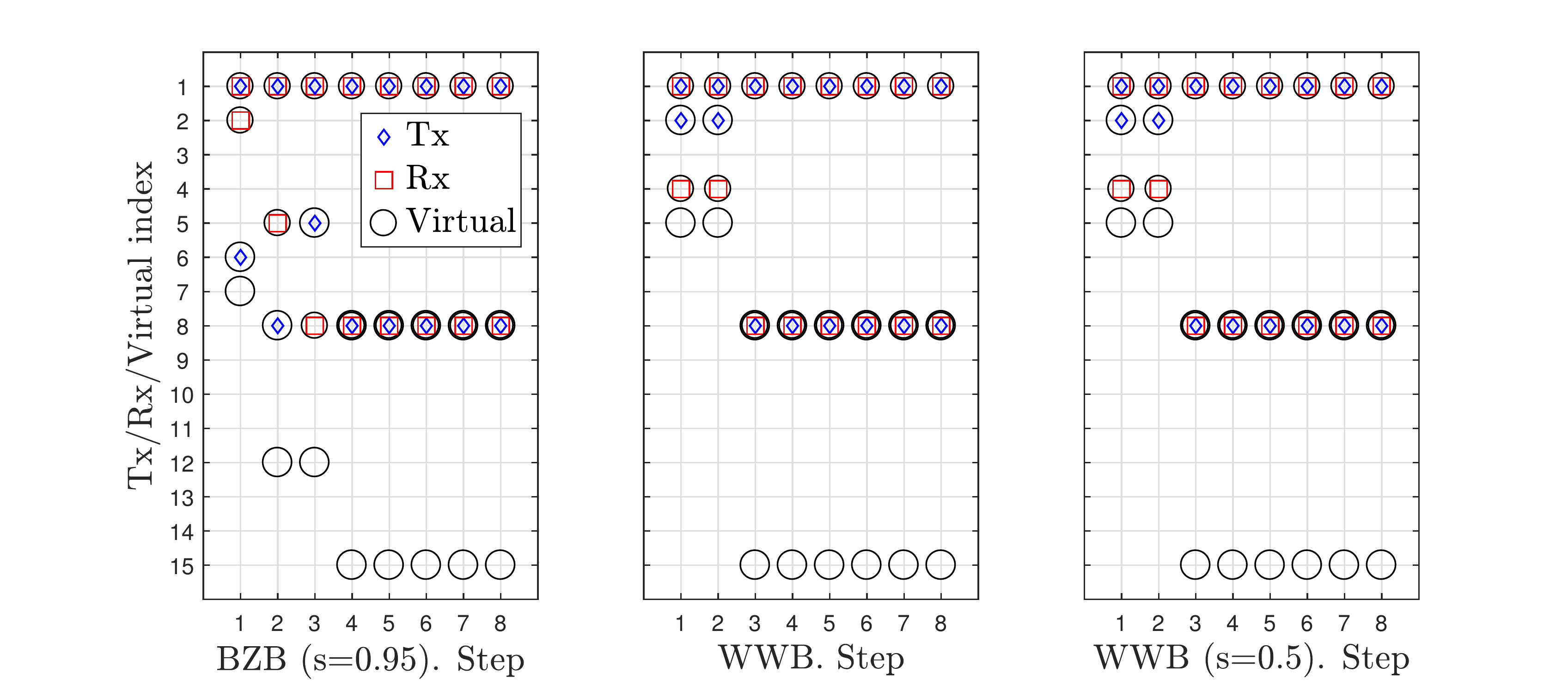}}
		\caption{Optimal channel choices in the MIMO case for each policy, under a Gaussian approximation of the posterior at each step, where Tx $1$ and Rx $1$ are always fixed. (Overlapping virtual elements are represented with concentric circles.)}
	\label{fig:channel-activations-MIMO}
		\end{center}
\end{figure}


\subsection{Simulations}


Here we compare in simulations the performance of channel selection policies that optimize the WWB, the BZB, and the ECRB\footnote{ 
	The optimization domain for the WWB is $(s,h)\in[.1,.9]\times[10^{-4}, 2]$; 
	for the BZB we use $s=0.95$, numerically more stable than $s=1$; and we use the ECRB instead of the BCRB because they yield equivalent policies.} for SIMO and MIMO arrays. 
%
The separation between adjacent transmitters and receivers is $0.9 \lambda/2$,
the number of snapshots is $J=2$,
the target signal~$\sjk$ is assumed known and equal to~$1$,
and we assume an initial prior distribution for the electronic azimuth uniform in $[-1, 1]$.
The target is static, $\fd=0$, and therefore the \textit{order} of transmitter activations is irrelevant for any given subset of them.
 We perform the inner optimization in~\eqref{eq:policy-optimization-wwb-framework} using simulated annealing~\cite{SK-CDG-MPV:83}\footnote{Matlab code, by H\'ector Corte, available in MathWorks File Exchange.}
 with a \textit{cooling speed} of $100$ \textit{intermediate} temperatures when the SNR is less than $0$, and $50$ otherwise,
  and the posterior is sequentially updated using a particle filter with residual resampling~\cite{RD-OC:05}\footnote{Matlab code ``Resampling methods for particle filtering,'' by J.-L. Blanco Claraco, 
 	available in MathWorks File Exchange.} and $500$ particles.
 

%

The channel choices for the SIMO and MIMO cases are shown in
Figs.~\ref{fig:channel-activations-SIMO} and~\ref{fig:channel-activations-MIMO}
for a single execution of our algorithm with
	 SNR$=-5$. These choices depend on the posterior distribution updated by each strategy and thus on the unique history of previous measurements and channel selections.
%
%
%
In the SIMO case, we observe a qualitative behavior for the policies that optimize the WWB and BZB analogous to the simulations in~\cite{OI-JT-IB:15,JT-OI-IB:16}, where during the first measurements receivers tend to be chosen closer together to avoid ambiguity in the estimation, and in subsequent measurements are selected farther apart to increase resolution. A similar behavior can be seen in the MIMO case.

\begin{figure}[h]
	\centering 
	{\includegraphics[width=0.99\linewidth]{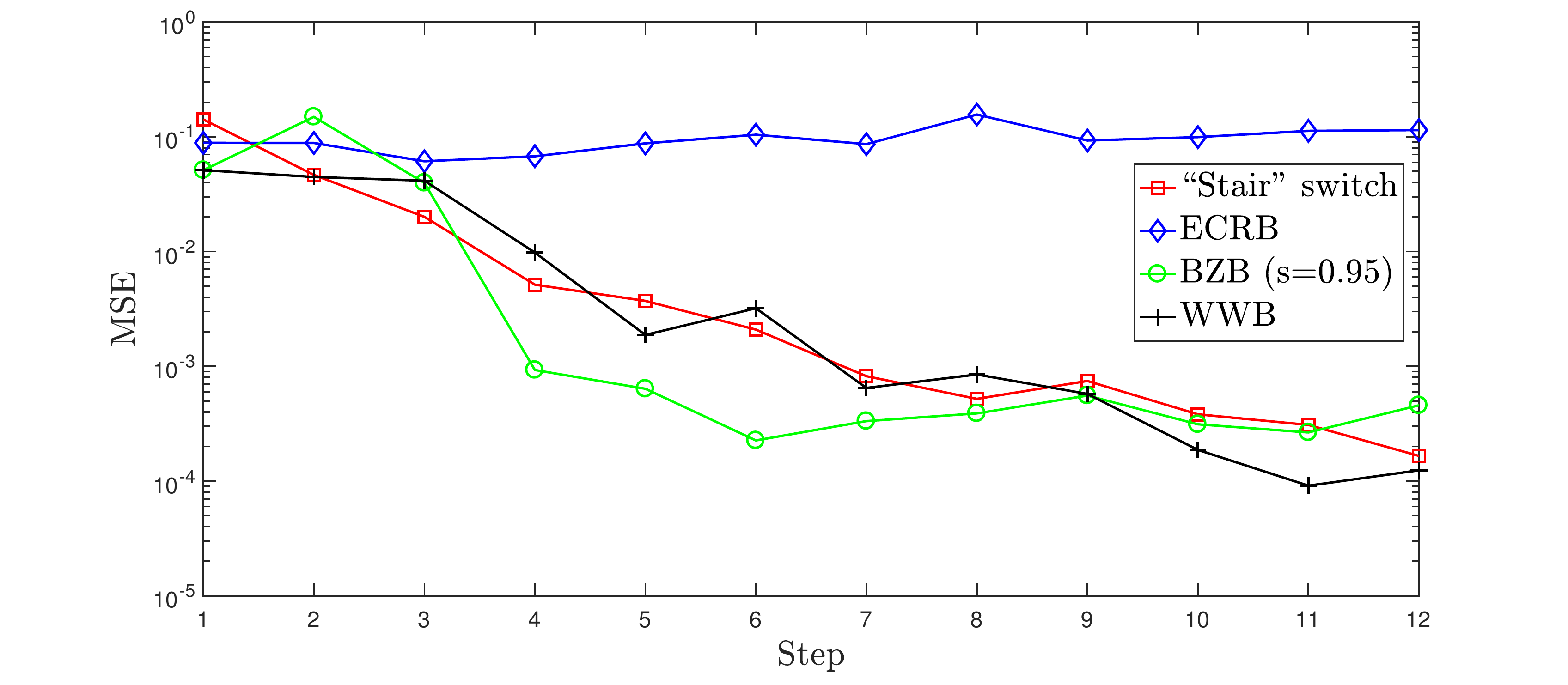}}

	{\includegraphics[width=0.99\linewidth]{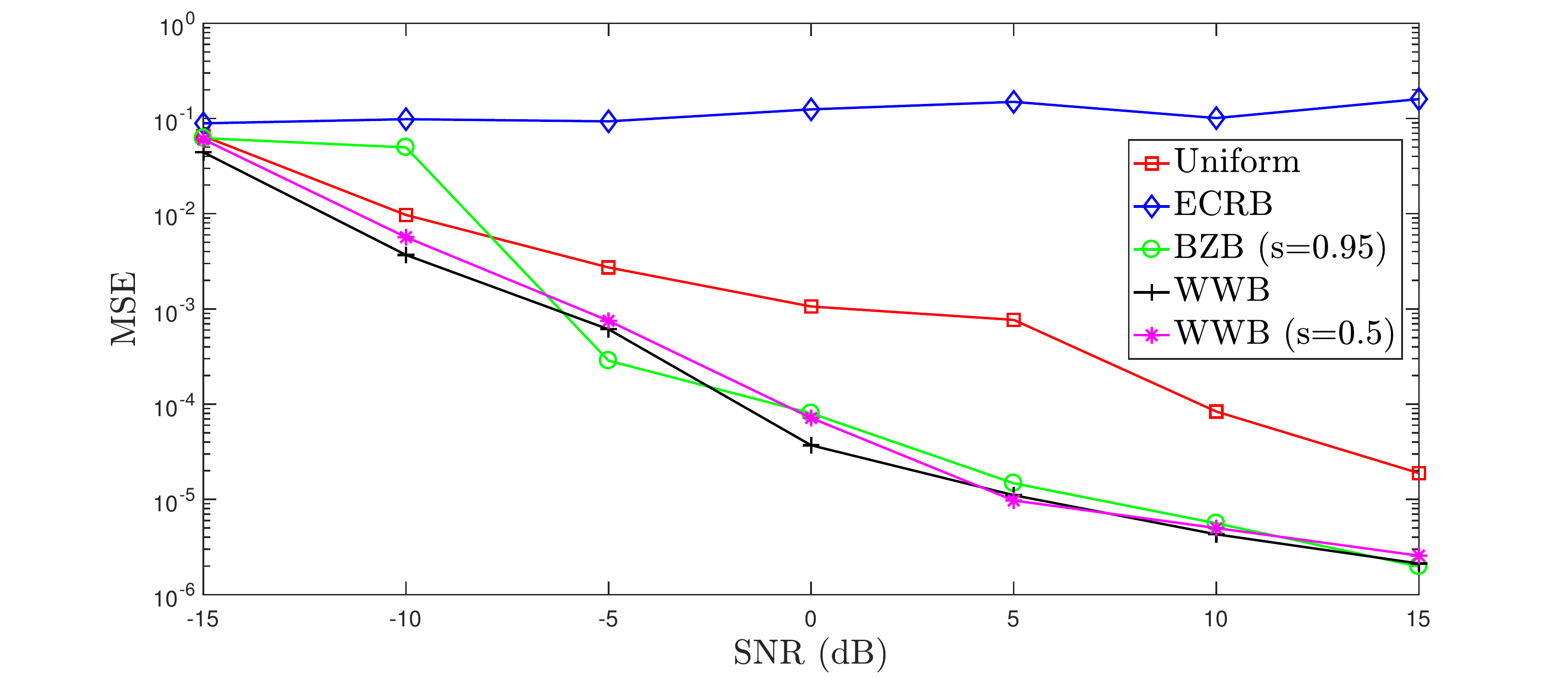}}
	
	\caption{MSE of the conditional mean for each policy. In the SIMO case (top) we depict a single execution over time with SNR $=-5$. In the MIMO case (bottom) we plot the average MSE at step~$8$, over $20$ executions, for each SNR. 
	}\label{fig:mse-comp-MIMO}
\end{figure}

We analyze the performance using the MSE of the conditional mean estimator~$\pargenh$ that results from each sensing policy. This is computed at each measurement step with respect to the true parameter value $\theta=\sin(\phi)=0.3$ using $300$ Monte Carlo realizations of each snapshot. In the SIMO case,
Fig.~\ref{fig:mse-comp-MIMO} (top) shows the same single execution as in Fig.~\ref{fig:channel-activations-SIMO}. 
In the MIMO case, we have simulated a computationally faster version of the adaptive policies where the expectation in~\eqref{eq:eta-posterior} is approximated replacing the posterior given by the particles by a Gaussian distribution with the same mean and variance. Using the result in~\cite[eqs. (138), (152)]{AR-PF-PL-CDR-AN:08}, this allows us to obtain a close form for~\eqref{eq:eta-posterior}.
%
  With this approximation, 
	Fig.~\ref{fig:mse-comp-MIMO} (bottom)
	compares the average MSE, over $20$ algorithm trajectories for each SNR, of the conditional mean estimator at measurement step~$8$. 	
%
%
%
%
We observe that optimizing the WWB yields slightly better performance than using the BZB for low SNR values.\footnote{Some preliminary analyses show this might be due to a ``more stable" behavior of the closed-loop that combines the optimal channel selection and the particle filter with regards to aliasing.}
In addition, these adaptive policies outperform \textit{ad hoc} strategies with the same number of active antennas, 
including the SIMO ``stair" switch and the fixed MIMO uniform virtual array. The evaluation of the WWB with $s= 0.5$ yields comparable performance to (but not always the same channel choices as) the WWB even though it uses a single test-point as the BZB. The discussion of the computational complexity depends on the cooling speed that is required for the optimization of each Bayesian bound and will be included in future work.



%
%
%

\section{Conclussions and future work}\label{sec:conclussions}
Adaptive strategies based on the Weiss-Weinstein bound outperform some common channel selections for DoA estimation.
The biggest concern is the computational time of policy evaluation at the controller, 
which for DoA estimation of a single target can be greatly reduced by fitting the output of the particle filter by a Gaussian,
and also the number of candidate subsets of channels.
 Future work also includes 
 target dynamics and estimation of  model parameters such as the SNR or the Doppler frequency, and employing multi-step ahead predictions.



\bibliographystyle{ieeetr} %

\end{document}